\documentstyle[12pt]{article}
\begin{document}
\begin{titlepage}
\pagestyle{empty}
\topmargin-.5in
\baselineskip=14pt
\rightline{UMN-TH-1240/94}
\rightline{CfPA-TH-94-13}
\rightline{LBL-35275}
\rightline{hep-ph/yymmddd}
\rightline{February 1994}
\vskip .2in
\baselineskip=18pt
\begin{center}
{\large{\bf On Preserving a $B+L$ Asymmetry Produced in the Early
Universe}}\footnote{This work was supported in part
by DoE grants
DE-FG02-94ER-40823 and DE-AC03-76SF00098,  by NSF grant AST-91-20005
and by NSERC.}
\end{center}

\vskip .1in
\begin{center}
Sacha Davidson

{\it Center for Particle Astrophysics, University of California}

{\it Berkeley, California, 94720}

Hitoshi Murayama\footnote{On leave of absence from {\it Department of
Physics, Tohoku University, Sendai, 980 Japan}}

{\it Theoretical Physics Group, Lawrence Berkeley Laboratory}

{\it University of California, Berkeley, CA 94720, USA}

and

Keith A. Olive

{\it School of Physics and Astronomy, University of Minnesota}

{\it Minneapolis, MN 55455, USA}

\vskip .1in

\end{center}
\vskip .5in
\centerline{ {\bf Abstract} }
\baselineskip=18pt
One of the most efficient mechanisms for producing the baryon asymmetry
of the Universe is the decay of scalar condensates in a SUSY GUT as was
first suggested
by Affleck and Dine. We show that given a large enough asymmetry, the baryon
number will be preserved down to low temperatures even if $B - L = 0$,
because the baryon number carrying scalars form bose condensates
that give the $W$  a mass.  We
derive
the conditions on the condensate needed to suppress electroweak
sphaleron interactions
which would otherwise drive the baryon asymmetry to zero when $B - L = 0$.

\noindent
\end{titlepage}
\baselineskip=18pt

\def\la{~\mbox{\raisebox{-.6ex}{$\stackrel{<}{\sim}$}}~}
\def\ga{~\mbox{\raisebox{-.6ex}{$\stackrel{>}{\sim}$}}~}
\def\b-l{$B - L$}
\def\bl{$B + L$~}
\def\tm{$\tilde m$}
\def\beq{\begin{equation}}
\def\eeq{\end{equation}}
\topmargin0.0in
\setcounter{footnote}{0}
Models for baryogenesis at the weak scale and above have been under
considerable
scrutiny since the realization that baryon number violation due to
non-perturbative
electroweak interactions is rapid at high temperatures \cite{vio}. Baryon
number violating
interactions mediated by sphalerons in fact violate \bl while
conserving \b-l. Thus any model of baryogenesis above the weak scale which
also conserves \b-l is likely to produce a negliglible baryon asymmetry
once sphaleron interactions have been incorporated \cite{am}.
This conclusion is clearly important as the simplest models
of baryogenesis typically conserve \b-l. There are, however, several
modifications
to these simplest scenarios which can overcome this problem.
One possibility is that the baryon asymmetry is generated during
the electroweak phase transition \cite{ckn}, though it is unlikely that this
can be done within the standard model. It is
possible that a more complicated grand unified
theory
 which violates \b-l, (such as $SO(10)$ rather than $SU(5)$)  produces a baryon
asymmetry which is only reshuffled
by sphaleron interactions. Or, an extension of the standard model
which violates lepton number and hence \b-l (for instance
 the inclusion of right handed neutrinos used to generate neutrino
masses via the
see-saw mechanism \cite{ygrs}), can produce
a lepton asymmetry which is transformed into a baryon asymmetry at the
weak scale \cite{fy1,cdo2,my}.

Alternatively, the conditions under which sphalerons are
able to destroy a prior
baryon asymmetry have come under consideration.
For example, lepton mass effects in the presence of  lepton
flavour asymmetries and $B+L$ violating electroweak interactions
can generate a baryon asymmetry even though $B+L = B-L = 0$ \cite{KRS,dr}. It
has also been shown that because
of the very small value of the electron Yukawa coupling constant, the
equilibrium condition
that leads to $B=L=0$ is only valid at late times (temperatures $T \la 1$ TeV)
 close to the electroweak
phase transition \cite{cdeo3,clkao}. At higher temperatures the decoupling
of the right-handed electron in some sense safe-guards the baryon asymmetry.
Though in the standard model sphalerons win in the end, the destruction of the
baryon asymmetry is
exponentially sensitive to parameters
of the model.   In an extension of the standard model the baryon
asymmetry may yet survive. There are also attempts to protect baryon
asymmetry introducing new fields \cite{ahr}.

In this letter, we explore another possibility. We consider the effect
of scalar Bose--Einstein (B-E)
condensates on the sphaleron interaction rate in a $B-L$ conserving
supersymmetric grand unified
theory. The presence of a B-E condensate gives rise to
gauge boson masses, which, if they persist down to temperatures
close to the electroweak
phase transition, can in principle suppress sphaleron interactions
sufficiently throughout the history of the universe
to protect the baryon asymmetry.

Let us first recall some facts about  baryogenesis in grand unified
theories. The simplest mechanism is the out-of-equilibrium decay of
heavy gauge or Higgs fields \cite{ww}. The same is
 generally true in a supersymmetric gut\footnote{See
Ref.~\cite{HMY} for a recent analysis in minimal SUSY $SU(5)$.}.  However,
 in the framework of
inflationary cosmology \cite{infl}, where generic models
 of inflation are constrained by
the observed anisotropies in the microwave background \cite{cobe},
 these out-of-equilibrium decay
scenarios are also constrained \cite{cdo2}.  In what follows
 below, we consider
only a generic inflationary model in which the duration of
 inflation and the magnitude
of density perturbations are determined by a single mass scale $\mu$.
We will assume that the inflaton potential can be written as
 $V( \psi ) = {{\mu}^4} P( \psi )$
where $\psi$ is the scalar field driving inflation, the inflaton,
and $P(\psi)$ is a
function of $\psi$ which possesses the features necessary for
inflation, but contains no small parameters.  The COBE data
fixes the mass scale $\mu$ relative to the Planck scale to be roughly
\begin{equation}
{\frac{\mu^2}{M_P^2} \simeq few\times{10^{-8}}}
\end{equation}

       Fixing $({\mu^2}/{M_P^2})$ has immediate general consequences
for inflation \cite{eeno,cdo2}. For example, the Hubble parameter during
inflation,
${{H^2} \simeq (8\pi/3)({\mu^4}/{M_P^2})}$ so that $H \sim
10^{-7}M_P$. The duration of inflation is $\tau \simeq
{M_P^3}/{\mu^4}$, and the number of e-foldings of expansion is $H\tau
\sim 8\pi({M_P^2}/{\mu^2}) \sim 10^{9}$. If the inflaton decay rate
goes as $\Gamma \sim {m_{\psi}^3}/{M_P^2} \sim {\mu^6}/{M_P^5}$, the
universe recovers at a temperature $T_R \sim (\Gamma{M_P})^{1/2} \sim
{\mu^3}/{M_P^2} \sim 10^{-11} {M_P} \sim 10^8 GeV$.

The relatively low inflaton mass $m_\psi \sim \mu^2/M_P \la 10^{12}$ GeV
is problematic for the out-of-equilibrium decay in a susy gut.  The low
mass scale would require at least one baryon number violating gauge or Higgs
boson
with a smaller mass causing proton decay at experimentally
disallowed rates.
There is a natural alternative to the out-of-equilibrium decay scenario in a
supersymmetric GUT, which is the decay of sfermion condensates as first
proposed
by Affleck and Dine \cite{ad}.  In this scenario, squark and slepton fields
obtain large vacuum expectation values along flat directions of the scalar
potential.  Effective baryon number violating operators
induce a baryon asymmetry in coherent flat direction oscillations.
The decay of these flat directions, due to supersymmetry breaking
effects characterized by a scale \tm, will produce a potentially large
baryon asymmetry if C and CP are violated (explicitly or
spontaneously) in that sector. If the oscillations
of the flat direction, $\phi$, come to dominate the energy density of the
Universe,
indeed a large baryon to entropy ratio, $n_B/s \sim O(1)$ is expected
\cite{ad,lin}.
It was already pointed out in the original paper that
B-E condensates of the squarks and sleptons may form, and
Dolgov and Kirilova \cite{dk} showed that the B-E condensate may
persist down to relatively low temperature in a Universe
where the energy density is dominated by the flat direction. We will show in
this letter that in an inflationary Universe,
the B-E condensate can persist down to the electroweak
phase transition
 for $n_B/s \sim O(10^{-2})$--$O(1)$. Then the sphaleron
transitions are effectively killed, and  the original
$B+L$ asymmetry generated by the decay of the sfermions along the flat
direction is preserved.

Note that this relatively large $n_B/s$ can be naturally obtained within the
inflationary cosmology and Affleck-Dine scenario.
In the context of inflation, it is quite likely that the Universe is dominated
by
the inflaton and the radiation products of inflaton decay rather than the
sfermion
oscillations.  Inflation, in fact, offers a natural explanation for the large
initial value of the sfermion fields. During inflation, quantum fluctuations
drive massless scalar fields as $\langle \phi^2 \rangle = H^3 t/ 4\pi^2
$ \cite{phi2}, which gives $\phi_o^2 \sim \mu^2 \sim m_\psi M_P$
at the end of inflation. In this case the baryon asymmetry produced is
somewhat lower as it is
diluted by the entropy produced by inflaton decay. The result is \cite{eeno}
\beq
\frac{n_{B}}{s} \simeq
\frac{\epsilon{\phi_o^4}{m_{\psi}^{3/2}}}{{M_X^2}
{M_P^{5/2}}{\tilde{m}}} \label{2}
\eeq
where $\epsilon$ is a combination of coupling constants whose
value
parametrizes the CP-violation, $\phi_o$
is the initial sfermion value, which is determined by quantum
fluctuations during inflation, and $M_X \simeq 10^{-4} - 10^{-3}
M_P$ is the unification scale.
For \tm~$\sim 10^2$--$10^3$ GeV, and
$m_\psi \sim 10^{11}$--$10^{12}$~GeV required from COBE data \cite{eeno,cdo2},
$n_B/s$ ranges from $10^{-6} \epsilon$
to $\epsilon$, making this scenario very attractive.
Note further that $\epsilon$ can be $O(1)$ since the  flat direction
vacuum expectation value (vev) can break CP spontaneously and the vev
 is the same over the whole universe due to  inflation.
So a relatively large value of $n_B/s$
 can be naturally obtained within
this scenario.

In a minimal SUSY GUT where \b-l is conserved, this asymmetry can be
destroyed by sphaleron interactions. (However, the Affleck-Dine scenario
itself does  extend to 
incorporate \b-l violation \cite{ceno,Morgan}). Here we are
interested in a \b-l
conserving theory, and we will see whether the sphaleron interactions
are indeed present when the asymmetry takes near maximal values. The sphaleron
interaction rate is roughly \cite{am}
\beq
\Gamma_{sph} \sim  { 10 ^{-2} m_W^7 \over \alpha_W^3 T^6}e^{-E_{sph}/T},
\label{sph}
\eeq
where  $E_{sph} \simeq 4 m_W/\alpha_W$ and the numerical value of the prefactor
depends on the gauge and higgs couplings. (Here we took $\lambda = g^2$
since this is a supersymmetric theory.)
In the presence of a condensate, $m_W \ne 0$\footnote{We refer the readers to
Refs.~\cite{Kapusta,HW,BBD} on discussions of the symmetry breakdown in the
presence of high number asymmetry.} and the sphaleron
rate is exponentially suppressed.  We will see that a condensate
is likely to be present at lower temperatures if the degeneracy due to a baryon
asymmetry is high,
 hence
we will assume $n_B/s \sim O(1)$, 
and discuss later how large $n_B/s$ needs to be.

In Ref.~\cite{eeno}, parameters were chosen so that $n_B/s$ was relatively
small,
{\it i.e.}\/ so that no additional entropy generation was required (more on
this below).
In that case a relatively simple sequence of events was found to occur:
Initially (after the exponential expansion due to inflation), the inflaton
begins
to oscillate at a value of the cosmological scale factor $R = R_\psi$.
Soon afterwards, the sfermions  start oscillating along the flat direction at
$R = R_\phi \simeq (m_\psi/{\tilde m})^{2/3} R_\psi$. The inflaton oscillations
decayed first at $R = R_{d\psi} = (M_P/m_\psi)^{4/3}R_\psi$ followed by the
decay of the
sfermion oscillations at $R = R_{d\phi} = (m_\psi^{7/15} \phi_o^{2/5}
M_P^{2/15}/
{\tilde m}) R_\psi$. For a given value of $R$, the number density of the
asymmetry can be written as
\beq
n_{B} = \epsilon {\phi_o^4 m_\psi^2 \over M_X^2
{\tilde m}} \left({R_\psi \over R}\right)^3 .  \label{5}
\eeq
At this point the radiation from inflaton decays has not yet
had enough time to thermalize, and equilibrium was established at the slightly
later time corresponding to $R= R_T \sim R_{d\psi}/\alpha^2$, where
$\alpha$ is a
gauge fine structure constant. The temperature of
the Universe now is well defined
and takes the value $T \sim m_\psi^{3/2}
\alpha^2 / M_P^{1/2} N_T^{1/3} \la 10^5$ GeV,
where $N_T \sim 200$ is the number of particle
degrees of freedom at $T$.

A large baryon asymmetry will make both
quantitative as well as qualitative changes
to this picture, because the presence of the flat direction or B-E
condensate gives  mass to the gauge and Higgs multiplets and the
interactions are suppressed as $\sigma \sim (\alpha^4/\pi^2) T^2
/\phi^4$ where $\phi$ is the typical amplitude of either the flat direction
or the B-E condensate. In particular, the history before  thermalization may
be drastically altered. However, it does not affect the baryon asymmetry,
 as we will see below.\footnote{We believe that the sphaleron
transitions are not present prior to  thermalization, since
collective excitations like the sphaleron can be formed only in a dense
plasma, while there is only a dilute gas of high energy particles before
 thermalization.}

Let us follow the evolution before thermalization for the sake of
completeness. For simplicity, we will
assume that the sfermion condensate decays
directly only to fermions.\footnote{This may happen if the
flat direction $\phi$ corresponds
to the lightest scalar field, making the decays
to scalars kinematically
forbidden. As we will see below, our results are independent of this
assumption.}
Prior to thermalization, chemical equilibrium between sfermions and
fermions is not reached,\footnote{This is true for the flat direction
in Ref.~\cite{Morgan} where all gauge interactions are killed by the
flat direction.} and
the decay to fermions
is Pauli blocked. Consider the decay
rate of a scalar $\phi$ to two massless fermions,
\beq
\Gamma_\phi = {1 \over 2{\tilde m}} \int {d^3p_1
\over (2\pi)^3 2E_1}{d^3p_2 \over (2\pi)^32E_2}
(2\pi)^4\delta^4(\Sigma p)|
{\cal M}|^2(1-f_1)(1-f_2)
\eeq
where $|{\cal M}|^2 = 2({\tilde m}^2/\phi^2)
F(g_i, h_j) (p_1 \cdot p_2)$ ; $p_1,p_2$
are the outgoing fermion four-momenta, and $F(g_i, h_j)$ is a function
of gauge and Yukawa coupling constants in both the numerator
and the denominator. We assume $F(g_i, h_j) \sim 1$. The factor $({\tilde
m}^2/\phi^2)$ accounts
for the suppression due to masses of the fields directly coupled to $\phi$.
 $f = (\exp((p-\mu)/T) + 1)^{-1}$ is the fermion
momentum distribution with chemical potential $\mu$.
At $T=0$, $f = 0$ for $E=p \ge \mu$
and $=1$ otherwise. Thus,
\beq
\Gamma_\phi = \left\{ \begin{array}{ll}
                       {{\tilde m}^3 \over 16\pi \phi^2} & \mu < {{\tilde m}
\over 2} \\
                          0 & \mu \ge {{\tilde m} \over 2}
\end{array} \right.
\eeq
 and the baryon density contained
in fermions is
\beq
n_B^{(f)} = {g_f \over 3 \pi^2} \mu^3
\eeq
where $g_f$ is the number of fermion species sharing the baryon asymmetry.
Clearly, $n_B^{(f)} \ll n_B$ for $\mu ={\tilde m}/2 $ since
$\tilde{m} \ll n_B^{1/3}$. The decay of the sfermion condensate
is therefore delayed.

We expect the Universe to thermalize at a similiar scale factor in the presence
of the condensate as in the model considered in \cite{eeno}.
Subsequent to thermalization, sfermion decays occur in a thermal bath, and the
decay to
fermions proceeds rapidly.

During the decay, equilibrium between fermions and scalars will be achieved and
the equilibrium condition $\mu_F= \mu_B = \tilde m$ will insure that the
baryon number flows back to scalars.
However, for a highly degenerate boson gas at low temperature (low with
respect to a critical temperature we derive shortly), a B-E
condensate will develop.  The total baryon asymmetry may then be written as
\beq
n_B = n_B^{(f)} + n_B^{(b)} + n_B^{(c)}  \label{10}
\eeq
where the superscripts denote contributions in fermions $(f)$,
($p \ne 0$) bosons $(b)$ and a B-E condensate $(c)$.
For $T \gg \mu = \tilde m$,\footnote{We treat the scalar particles as
free bose gas. Then $\mu = \tilde{m}$ is the maximum possible chemical
potential.}
it is straightforward to expand in $\mu/T$, neglecting
${\tilde m}$, to compute $n_B^{(f)}$ and $ n_B^{(b)}$
\begin{eqnarray}
n_B^{(f)} &=& {g_f \over 6} \mu_F T^2  \label{11}\\
n_B^{(b)} &=& {g_b \over 3} \mu_B T^2  \label{12}.
\end{eqnarray}
We
can define a (scale factor dependent) critical temperature $T_c$
below which a B-E condensate will be present
 by
\beq
 n_B^{(b)} + n_B^{(c)} = {g_b \over 3} {\tilde m} T_c^2
\label{tc}
\eeq
If the temperature of the Universe is below $T_c$, the baryon number
in the B-E condensate will be
\beq
n_B^{(c)} = {g_b \over 3} {\tilde m}
 \left(1 - \left({T \over T_c}\right)^2 \right) T_c^2 \label{14}
\eeq

So far we have followed the history until  thermalization.
Now we follow the evolution after the thermalization. At $R > R_T$,when the
decay is complete, and assuming equilibration,
$n_B^{(f)} \ll n_B$ (since $\mu T^2 \ll T^3 \sim n_B$) and $n_B^{(b)} = 2
n_B^{(f)}$ so the baryon number is now almost entirely in the form of the B-E
condensate.  One sees now that the assumption of the sfermion
decay to fermions is not crucial as decay into ($p \ne 0$) bosons would not be
possible
if $T$ is below $T_c$. The critical temperature evaluated at $R_T$ is
(from eqns (\ref{tc}) and (\ref{5}) with $\epsilon = 1$) $T_c
\simeq (3 n_B/g_b {\tilde m})^{1/2} \sim (3/g_b)10^7$
GeV. Notice that the presence of the B-E condensate at this late stage
($T \sim 10^5$ GeV) depends crucially on a large baryon asymmetry.
{}From eq. (\ref{tc}) one sees that while $T_c \sim R^{-3/2}$ (because
$n_B$ scales like $R^{-3}$),
$T \sim R^{-1}$,  so as the Universe expands
and cools, it will eventually  be at a temperature
above the critical temperature, and the B-E condensate will
evaporate. The phase transition will occur at a scale
$R_F \sim 10^3 R_T$ or at a temperature $T_F \sim 100$ GeV, {\em below}
the normal electroweak phase transition temperature $\equiv T_{EPT}$.
(The presence of the B-E condensate could affect the electroweak phase
transition, because $SU(2)$ is broken above the EPT, and the scalar
vacuum expectation value (vev)
will give masses to most particles. However, if the B-E condensate
evaporation temperature $T_c$ is below $T_{EPT}$, it should be safe
to assume that there is always a vev present.)

We would like to estimate how large a baryon to entropy ratio
is required to make this
scenario work. Although
the Affleck-Dine flat direction consists of vevs for only
a small number of fields, we expect the equilibrium B-E condensate at low
temperatures to share the baryon number equally among all the squarks
(assuming that they all have similiar masses), and the lepton number
$L_i$ equally among the $i$th generation sleptons. For convenience,
let us assume that all the lepton number is in muons (as it would be in
Affleck and Dine's  flat direction), in which case
the slepton vevs will be slightly larger than those of the squarks,
because there are fewer of them among whom to share the
asymmetry. We require that  the  three B-E-condensed
scalars carrying muon number remain as
large as $\langle\phi_i\rangle \sim T/3$
at $T_{EPT} \sim$ 300 GeV (more later on this choice of
$\langle\phi_i\rangle$),
so that
\beq
n_{\mu} \simeq \tilde{m} T_{EPT}^2 + \frac{1}{2} \tilde{m} T_{EPT}^2 + 3
\tilde{m} \frac{T_{EPT}^2}{9}
\eeq
Taking $s = 2 \pi^2 N T^2/45$ gives
  \beq
\frac{n_{B}}{s} \simeq \frac{ n_{\mu}}{s}~\ga~\frac{45 \tilde{m}}
{\pi^2 N_{T_{EPT}} T_{EPT}} \simeq 0.01
\eeq
where we have used $\tilde{m} = 100$ GeV\footnote{$\tilde{m}$ here is
the slepton mass, and not necessarily equal to $\tilde{m}$ of equation
(\ref{2}), which is the overall supersymmetry-breaking mass.}, $T_{EPT} = 300$
GeV,
and $N_{T_{EPT}} = 200$.
If we assume that there is no entropy generation between
$R_{EPT}$ and $R_T$, then to preserve a B-E condensate down to
below the EPT we need $n_{B}/s \ga 0.01$. As we discussed
previously, this should be possible.

We have shown that given a large enough baryon asymmetry,
a scalar condensate will
persist down to very low temperatures.  This same
conclusion was found in another
context by Dolgov and Kirilova \cite{dk}.
However in order to protect the baryon
asymmetry against wash-out from sphaleron interactions, we need to still show
that the sphaleron interaction rate  (eq. \ref{sph})  is supressed.
  If we define $v = \sqrt{2\sum_i |\phi_i|^2}$,
where $\phi_i$ are the B-E condensate vevs carrying $SU(2)$ quantum numbers,
thus $m_W = gv/2$, and
\beq
{\Gamma_{sph} \over H} \simeq \frac{g v^7 M_P}{12 N_T^{1/2} T^8}
\exp\left[-\frac{8 \pi v}{g T}\right]
\eeq
which is  out-of-equilibrium at $T \simeq 300$ GeV for $v \ga .8 T$.
So if we assume that the vevs corresponding to the lepton condensates
are larger than $T/3$ near the electroweak phase transition, as we
required in the previous paragraph, the sphalerons are clearly out
of equilibrium.

Finally, we wish to comment on the problem of entropy production.
Typically this is usually a problem concerning excess entropy.  Here,
we are in a perhaps more enviable situation requiring some entropy
generation at or around the weak scale.  There has been a considerable
amount of discussion of excess entropy production in supergravity
and superstring-inspired models \cite{ent}.  This question was addressed
specifically in the context of the AD mechanism for baryogenesis in
\cite{eeno1}.
There a constraint on intermediate scale models was derived
to avoid the overproduction of entropy.  A relaxation of that bound
could possibly yield the entropy needed here.\footnote{Using the
dilution factor
$\Delta \sim 10^2 m_I^3/\tilde{m}^{5/2} M_P^{1/2}$ (Eq.~(14) in
\cite{eeno1}), one indeed obtains the desired $\Delta \sim 10^7$ for
$m_I \sim 10^7$~GeV, corresponding to $n=1$ in their Eq.~(6).}

In summary, we have  shown that if the baryon to entropy ratio
at thermalisation is large ($\ga 0.01$) in a theory with $B-L=0$,  $B+L$
carrying bose condensates may keep the sphalerons out of equilibrium
until the electroweak phase transition, and thereby prevent them from
washing out the asymmetry.

\vskip 0.8truecm
\noindent {\bf Acknowledgements}
\vskip 0.4truecm
We would like to thank Bruce Campbell and John March-Russell for useful
discussions.
This work was supported in part by the Director, Office of Energy
Research, Office of High Energy and Nuclear Physics, Division of High
Energy Physics of the U.S. Department of Energy under Contracts
DE-FG02-94ER-40823 and
DE-AC03-76SF00098, by NSF grant AST-91-20005 and by NSERC.
The work of KAO was in addition supported by a Presidential Young
Investigator Award.

\end{document}